\documentclass[12pt]{article}
\usepackage{epsf}
\usepackage{graphicx,graphics,epsfig}
\usepackage{amsmath,amssymb,graphicx,lscape,txfonts}
\usepackage{rotating}
\usepackage{xcolor}
\def\Cov{\mbox{Cov}}

\def\E{{E}}

\def\sgn{\mbox{sgn}}
\begin{document}
\baselineskip 8mm
\setcounter{page}{0}
\thispagestyle{empty}
\begin{center}
{\Large \bf An efficient Gehan-type estimation  for the accelerated
failure time model with clustered  and censored data} \vspace{3mm}

Liya Fu$^{a}$, Zhuoran Yang$^{a}$, Yan Zhou$^{b,*}$ and You-Gan Wang$^{c}$

{\it $^a$School of Mathematics and Statistics, Xi'an Jiaotong University, China\\
$^b$ College of Mathematics and Statistics, Institute of Statistical Sciences, Shenzhen University, China \\
$^c$ School of Mathematical Sciences, Queensland University of Technology, Australia}  \\
$^*${\it Email:zhouy1016@szu.edu.cn}

\end{center}

\centerline{\bf  ABSTRACT}

In medical studies, the collected covariates usually contain
underlying  outliers. For clustered/longitudinal data with censored
observations, the traditional Gehan-type estimator is robust to
outliers existing in response but sensitive to outliers in the
covariate domain, and it also ignores the within-cluster correlations.
To take account of within-cluster correlations, varying cluster
sizes, and outliers in covariates, we propose  weighted Gehan-type
estimating functions for  parameter estimation in the accelerated
failure time model for clustered data. We provide the asymptotic
properties of the resulting estimators and carry out
simulation studies to evaluate the performance of the proposed
method under a variety of realistic settings.
The simulation results demonstrate that the proposed method
is  robust to the outliers existing in the covariate domain and lead
to much more efficient estimators when a strong within-cluster
correlation exists.  Finally, the proposed method is applied to a
medical dataset and more reliable and convincing results are
hence obtained.


\newpage

\section{ Introduction}

Censored  data are very common in biomedical studies.
A popular method for analyzing censored data is the Cox proportional
hazards model (Cox 1972). However, when the proportional hazards
assumption is violated,  the Cox model may derive inconsistent
parameter estimators. A  semiparametric accelerated failure time
model is an alternative to the  Cox proportional hazards model,
which is a linear model for  the logarithm of the failure time and
covariates with error distribution unspecified (Kalbfleish and
Prentice 2002). Rank-based estimation for the accelerated failure
time model with clustered/longitudinal data has been studied by some
researchers in recent years (Lee et al. 1993; Lin et al. 1998; Jin
et al. 2006; Wang and Fu 2011; Chiou et al. 2015; Chiou et al.
2014). The analysis of the clustered failure time data is much more
complicated due to the potential  within-cluster correlations and
the nature of censoring. Lee et al. (1993) studied the weighted
log-rank statistic and the Buckley-James method for the correlated
and censored data, and provided the covariance matrix estimation of
the estimating functions, which does not require specifying the
error distribution. Jin et al. (2006) proposed rank-based estimating
functions for multivariate failure time data and developed a novel
resampling method for the covariance matrix estimation of regression
parameters. Chiou et al. (2015) presented weighted rank-based
estimating equations for fitting the AFT model with clustered
failure times from stratified random sampling, and used the induced
smoothing approach proposed by Brown and Wang (2005) to reduce the
computational burden. This approach has been adapted to clustered
failure time data by Johnson and Strawderman (2009) and Wang and Fu
(2011).

The aforementioned methods are based on the independence working
model assumption and ignore the underlying within-cluster
correlations. To take account of the within-cluster correlations and
improve the efficiency of estimators with similar computational
complexity for clustered survival data analysis, Wang and Fu (2011)
proposed splitting the Gehan weight estimating function to the
between- and within-cluster estimating functions and recombining the
two resultant estimators. Chiou et al. (2014) extended the
generalized estimating equations approach to the clustered and
censored data. In longitudinal studies, some potential outliers
exist in response and/or covariates, which often result in serious
problems for parameter estimation in the AFT model. The rank-based
method is robust against outliers in response. However, as far as we
know, the literature on parameter estimators against outliers in
covariates for clustered and censored data is quite limited. Luo et
al. (2014) proposed robust approaches based on the smoothed Gehan
rank estimation methods, but their method was based on an
independence model. This leads us to seek an efficient and doubly
robust method for clustered and censored data with outliers in
covariates and/or response.

In this paper, we propose weighted Gehan-type estimating functions
with the induced smoothing approach, which take account of the
within-cluster correlations, varying cluster sizes, and outliers in
covariates and/or response. Therefore, the proposed method is robust
against outliers existing in covariates and/or response.
Furthermore, the induced smoothing method is utilized to eliminate
computational issues resulting from the
 unsmoothness of the estimating functions and multiple solutions.
The induced estimating functions are continuous and differentiable,
which make the statistical inference convenient and provide both
regression parameter estimation and their covariance matrices. The
asymptotic properties of the estimators from the nonsmoothed
weighted rank-based estimating functions  are established. The
estimators from the smoothed estimating functions are shown to be
consistent and have the same asymptotic distribution as those from
the nonsmooth version. The covariance of the estimators is estimated
by a  sandwich formula.

In Section 2, we briefly review the accelerated failure time model,
and present the weighted rank-based estimating equations for the AFT
models with clustered data. In Section 3, we  provide computational
procedures for computing the parameter estimates and their
covariance and  carry out simulation studies to evaluate the
performance of the proposed method. In Section 4, we analyze two
real medical datasets for illustration. Some conclusions are
summarized in Section 5.

\section{ Weighted estimating functions }

\subsection{ The AFT model}

Suppose that there are $N$ independent clusters, and
their respective cluster sizes are $n_1,\cdots,n_N$. Let $T_{ik}$
and $C_{ik}$ denote the failure time and censoring time for the
$k$th member of the $i$th cluster, and let $X_{ik}$ be the
corresponding $p \times 1$ vector of covariates. We assume that
$(T_{i1}, \cdots, T_{i n_i} )'$ and $(C_{i1},\cdots, C_{i n_i} )'$
are independent conditional on the covariates $(X_{i1}, \cdots, X_{i
n_i} )'$. The accelerated failure time model is
\begin{eqnarray}\label{eq:model}
\log T_{ik}=X_{ik}^{\rm T}\beta+\epsilon_{ik}, \  \ k=1,\cdots,
n_i;\ \ i=1,\cdots ,N,
\end{eqnarray}
where $\beta$ is the unknown regression parameter vector
corresponding to the covariate vector $X_{ik}$ of dimension $p$, and
$(\epsilon_{i1}, \cdots, \epsilon_{ik_i} )'  $ are independent
random error vectors for $i = 1, \cdots, N$. However, for each
cluster, the error terms $\epsilon_{i1}, \cdots, \epsilon_{in_i}$
may be  correlated. If $\tilde{T}_{ik}=T_{ik}\wedge C_{ik}$, and
$\Delta_{ik}=I(T_{ik}\leq C_{ik})$, where $I(\cdot)$ is the
indicator function, then the observations consist of
$(\tilde{T}_{ik},\Delta_{ik},X_{ik})$.

\subsection{ Weighted estimating functions for AFT models}

Let $e_{ik}=\log\tilde T_{ik}-X_{ik}^{\rm T}\beta$.  The rank-based
estimating functions of dimension $p$ using the Gehan-type weight take the following form,
\begin{eqnarray}\label{gehan}
S_{G}(\beta)=N^{-2}\sum_{i=1}^N \sum_{j=1}^N \sum_{k=1}^{n_i}
  \sum_{l=1}^{n_j}\Delta_{ik}(X_{ik}-X_{jl})I(e_{ik}-e_{jl}\leq 0),
\end{eqnarray}
which is monotonic with respect to $\beta$ (Fygenson and Ritov
1994). Let $\hat\beta_G$ be the resultant estimator from
(\ref{gehan}), which can be also obtained by minimizing the
following scalar objective function,
\begin{eqnarray*}
L_{G}(\beta) &=&N^{-2}\sum_{i=1}^N\sum_{j=1}^N \sum_{k=1}^{n_i}
\sum_{l=1}^{n_j}\Delta_{ik}(e_{ik}-e_{jl})^{-},
\end{eqnarray*}
where $e^{-}=|e|I(e<0)$.

Because $S_G(\beta)$ is based on the independent working correlation
assumption, the efficiency of $\hat\beta_G$
 can be enhanced by accounting for the within-cluster correlations and the
impacts of varying cluster sizes. Furthermore, $\hat\beta_G$ is
robust against outliers in response and is sensitive to outliers in
covariates. To seek  doubly robust and efficient parameter
estimates, we propose the following weighted estimating functions
\begin{eqnarray*}
S_{\omega
h}(\beta)=N^{-2}\sum_{i=1}^N\sum_{j=1}^N\sum_{k=1}^{n_i}
\sum_{l=1}^{n_j}\omega_i\omega_jh_{ik}h_{jl}\Delta_{ik}(X_{ik}-X_{jl})I(e_{jl}<
e_{ik}),
\end{eqnarray*}
where $\omega_i$ and $h_{ik}$ are weights to be specified.
Let $\hat\beta_{\omega h}$ be the  estimator from $S_{\omega
h}(\beta)=0$, which can be also derived by minimizing the following
objective function
\begin{eqnarray*}
L_{\omega h}(\beta) &=&N^{-2}\sum_{i=1}^N\sum_{j=1}^N
\sum_{k=1}^{n_i}
\sum_{l=1}^{n_j}\omega_i\omega_jh_{ik}h_{jl}\Delta_{ik}(e_{ik}-e_{jl})^{-}.
\end{eqnarray*}

For  $\omega_i$, in a general way, we can select weights including
$\omega_i= 1$,  $\omega_i = 1/n_i$, and  $\omega_i = \{1 + (n_i -
1)\bar{\rho}\}^{-1}$, where $\bar{\rho}$ is the average
within-cluster correlation and is obtained using the moment
estimator from Wang and Carey (2003) given a consistent estimation
for $\beta$,
$$\hat{\bar{\rho}}=\frac{\sum_{i=1}^N\sum_{j\neq l}^{n_i} (r_{ij}-\bar{r})(r_{il}-\bar{r})}{\sum_{i=1}^N(n_i-1)\sum_{j=1}^{n_i}(r_{ij}-\bar{r})^2},$$
where $r_{ij}$ is the rank of the corresponding residual term
$e_{ij}$, and $\bar{r} = (M+1)/2$ is the average of the rank sum of
all $\{e_{ik}\}$. In this paper, we use the third weight to
incorporate the within-cluster correlations. For weight $h_{ik}$, we
use the generalized rank (GR) estimation (Naranjo and Hettmansperger
1994) defined by
$$h_{ik}=\min \left\{1,\left[\frac{c}{d^2_i(X_{ik})}\right]^{\alpha/2}\right\}.$$
 Here, $c$ and $\alpha$ correspond to tuning constants and $d^2_i(X_{ik})$
 denotes the squared Mahalanobis
distance of $X_{ik}$ based on the robust estimates of location and
dispersion for the design set $\{X_{ik}\}$ (Rousseeuw and van
Zomeren 1990). For the tuning parameters $\alpha$ and $c$, we use
$\alpha=2$ and $c=\chi_{0.95}^2(p)$ which is the $95$th percentile
of a $\chi^2(p)$-distribution. Specifically, when
$\omega_i=\omega_j= 1$ and $h_{ik}=h_{jl}=1$, $S_{\omega h}(\beta)$
corresponds to the classical Gehan-type estimating function
$S_G(\beta)$.

Denote $\beta_0$ as the true value of $\beta$. According to Lee et
al. (1993) and Jin et al. (2006), under some regularity conditions,
the limiting distribution of $\sqrt{N}(\hat\beta_{\omega
h}-\beta_0)$ follows a zero-mean multivariate normal distribution,
and the asymptotic covariance matrix of $\sqrt{N}\hat\beta_{\omega
h}$ is
\begin{equation}\label{eq:sigmaW}
 \Sigma_{\omega h}=\{D_{\omega h}(\beta_0)\}^{-1}V_{\omega h}\{D_{\omega h}(\beta_0)\}^{-1},
\end{equation}
where  $D_{\omega h}(\beta)=\partial{E(S_{\omega
h}(\beta))}/\partial{\beta^{\rm T}}$ and $V_{\omega
h}=\lim_{N\rightarrow \infty}\Cov\{\sqrt{N}S_{\omega h}(\beta_0)\}$.
 According to Lee et al. (1993), we deduce the
limiting variance matrix of $\sqrt{N}S_{\omega h}(\beta_0)$ given as
follows:
$$\hat V_{\omega h}=\frac{1}{N}\sum_{i=1}^N\sum_{k=1}^{n_i}\sum_{l=1}^{n_i}\omega_i^2h_{ik}h_{il} \hat\xi_{ik}(\beta)\hat\xi_{il}^{\rm
T}(\beta),$$
 where
\begin{eqnarray*}
\tiny\hat{\xi}_{ik}(\beta)&=&\sum_{j=1}^N\sum_{f=1}^{n_j}\omega_j\left\{\frac{\Delta_{ik}}{N}h_{jf}(X_{ik}-X_{jf})I(e_{ik}<
e_{jf})
 - \frac{\Delta_{jf}}{N}z_{ikrs}I(e_{ik}\geq
 e_{jf})\right\},
\end{eqnarray*}
\begin{eqnarray*}
z_{ikrs}=\frac{\sum_{r=1}^N\sum_{s=1}^{n_r}\omega_rh_{rs}(X_{ik}-X_{rs})I(e_{rs}\geq e_{jf})}{\sum_{m=1}^N\sum_{t=1}^{n_m}I(e_{mt}\geq e_{jf})}.
\end{eqnarray*}

Matrix $D_{\omega h}(\beta)$ depends on the error distributions,
which are unknown and usually difficult to estimate. If $S_{\omega
h}(\beta)$ is a  smooth function of $\beta$, $D_{\omega h}(\beta)$
can be estimated by $\partial{D_{\omega
h}(\beta)}/\partial{\beta^{\rm T}}$ evaluated at an estimate of
$\beta$. However, $S_{\omega h}(\beta)$ is a step function, Hence,
its derivative does not exist, which makes parameter estimates and
computation of $\Sigma_{\omega h}$ difficult. Moreover,
calculational issues often arise when minimizing $L_{\omega
h}(\beta)$ or solving $S_{\omega h}(\beta)=0$, and the solution in
general is not unique and consists of a single interval or even
multiple intervals, although the length of these intervals converges
asymptotically to zero (Jin et al. 2006).

\subsection{ Smoothed weighted estimating function}

To  overcome difficulties with the lack of smoothness of the
estimating functions, we now introduce the induced smoothing method
given by Brown and Wang (2005). Assume that $Z\sim N(0, I_p)$ and is
independent of the data, where $I_p$ denotes the $p \times p$
identity matrix. Let $\Gamma$ be a $p \times p$ positive definite
matrix and satisfy $||\Gamma||=O(1)$. Then, the induced smoothing
version of
 $S_{\omega h}(\beta)$ is $\tilde S(\beta) = {\rm E}_Z \{S_{\omega h}(\beta+ N^{-1/2}\Gamma Z)\}$,
where the expectation is taken with respect to $Z$.
 By some simple
calculation, we have
\begin{equation}
\tilde{S}_{\omega
h}(\beta)=N^{-2}\sum_{i=1}^N\sum_{j=1}^N\sum_{k=1}^{n_i}\sum_{l=1}^{n_j}\omega_i\omega_jh_{ik}h_{jl}\Delta_{ik}(X_{ik}-
X_{jl})\Phi\left(\frac{N^{1/2}(e_{jl}- e_{ik})}{r_{ikjl}}\right),
\label{eq:jy3}
\end{equation}
where $r^2_{ikjl}=(X_{ik}-X_{jl} )^T\Gamma^2(X_{ik}-X_{jl})$.  Let
$\phi(\cdot)$ be the standard normal density function. Similarly, we
can obtain the induced smoothing version of
 $L_{\omega h}(\beta)$,
\begin{equation*}
\tilde{L}_{\omega
h}(\beta)=N^{-2}\sum_{i=1}^N\sum_{j=1}^N\sum_{k=1}^{n_i}\sum_{l=1}^{n_j}\omega_i\omega_jh_{ik}h_{jl}\Delta_{ik}\left[(e_{jl}-e_{ik})
\Phi\left(\frac{N^{1/2}(e_{jl}-
e_{ik})}{r_{ikjl}}\right)+\frac{r_{ikjl}}{N^{1/2}}\phi\left(\frac{N^{1/2}(e_{jl}-
e_{ik})}{r_{ikjl}}\right) \right].
\end{equation*}
Then we can obtain $\tilde{\beta}$ by minimizing $\tilde{L}_{\omega
h}(\beta)$. Alternatively,  $\tilde{\beta}_{\omega h}$ can be
derived as the multivariate root of $\tilde S_{\omega h}(\beta)=0$.
The derivative of $\tilde S_{\omega h}(\beta)$ can be easily
derived,
\begin{eqnarray*}
\tilde D_{\omega h}(\beta) & = &  N^{-2}\sum_{i=1}^N \sum_{j=1}^N
\sum_{k=1}^{n_i}
\sum_{l=1}^{n_j}\omega_i\omega_jh_{ik}h_{jl}\Delta_{ik}
                     \frac{(X_{ik}-X_{jl})^T(X_{ik}-X_{jl})}{r_{ikjl}}\phi\left(\frac{N^{1/2}(e_{jl}-e_{ik})}{r_{ikjl}}\right) .
\end{eqnarray*}
 Before giving the  asymptotic properties of the smoothed
versions and the resultant estimators, the following regularity
conditions
are required.\\
C1. The parameter vector $\beta$ lies in a compact subset $\mathbb{B}$ of $\mathbb{R}^{p}$.\\
C2. For $i=1,\cdots, N$, $n_i$ are bounded and
    $\max_{1\leq k \leq n_i, 1\leq i \leq N}||X_{ik}||^2=o(\sqrt{N})$, where $ ||\cdot||$ is the Euclidean norm.\\
C3. ${\rm E}(\epsilon_{ik}^2)\leq M<\infty$, where $M$ is a constant. \\
C4. The common marginal density function of the errors
$\epsilon_{ik}$, $f_{0}(\cdot)$, and its derivative
$f_0'(\cdot)$ are bounded and satisfy $\int[f'_0(t)/f_0(t)]^2f_0(t)\mathrm{d}t<\infty.$\\
C5. The marginal distribution of $C_{ik}$ is absolutely continuous
    and the corresponding density function is bounded on $\mathbb{R}$, for any $i$ and
    $k$.\\
C6. The matrices $D_{\omega h}(\beta)$ and $\tilde D_{\omega
h}(\beta)$ are non-singular.

{\bf Theorem 1.}{\it Under conditions C1-C5,
$\sqrt{N}\{\tilde S_{\omega h}(\beta)-S_{\omega h}(\beta)\}=o_p(1)$
uniformly in $\beta$.}

{\bf Theorem 2.}{\it Let $\Gamma^2$ be any symmetric and positive definite
matrix with $\|\Gamma\|<\infty$. Under conditions C1-C6,
$\tilde{\beta}_{\omega h}$ is a strongly consistent estimator of
$\beta_0$.}

{\bf Theorem 3.}{\it Let $\Gamma^2$  be any symmetric and positive
definite matrix with $\|\Gamma\|<\infty$. Under conditions C1-C6,
$N^{1/2}(\tilde{\beta}_{\omega h}-\beta_0)$ converges in
distribution to $N(0, \Sigma_{\omega h})$, where $\Sigma_{\omega h}$
is given by (\ref{eq:sigmaW}).}

 Theorem 1 indicates that the difference between the
smoothed estimating functions and unsmoothed version is negligible.
Theorems 1 and 2 indicate that $\tilde D_{\omega h}$ is a consistent
estimator of $D_{\omega h}$. Therefore, $\Sigma_{\omega h}$ can be
consistently estimated using $\hat\Sigma_{\omega
h}=\tilde{D}_{\omega h}^{-1}(\beta)\hat{V}_{\omega
h}\tilde{D}_{\omega h}^{-1}(\beta).$ The proofs of theorems 1 and 2
are given in the Appendices A and
 B. The proofs of Theorem 3
can be established by following similar lines as in established
similar results for the independence estimator (Johnson and
Strawderman 2009).
According to Brown and Wang (2005), an iteration
procedure to simultaneously obtain the smoothed estimate
$\tilde{\beta}_{\omega h}$  and
its covariance matrix estimate can be described by the following  steps:\\
Step 1. Choose an initial value (e.g. $I_p$) for the working covariance matrix ${\Gamma}^{(0)}$ and a consistent estimator for $\beta$ to evaluate $w_i$ and $w_j$.\\
Step 2. In the $k$-th iteration, update $\tilde\beta_{\omega
h}^{(k)}$
        by minimizing $\tilde L_{\omega h}(\beta)$  or solving $\tilde S_{\omega h}(\beta)=0$.\\
Step 3. Update $\Gamma^{(k)}$ based on $ \Gamma^2=\{\tilde D_{\omega
h}(\beta)\}^{-1}\hat{V}_{\omega h}\{\tilde D_{\omega
h}(\beta)\}^{-1}$ using the current values $\tilde\beta_{\omega
h}^{(k)}$  and $\Gamma^{(k-1)}$.
        \\
Step 4. Repeat Steps 2-3 until a convergence criterion is satisfied.

In our experience, in general, the algorithm converges after only a
few iterations. The final values of $\tilde\beta_{\omega h}$ and
$\Gamma^2$ can be as estimates of $\beta$ and $\Sigma_{\omega h}$.

\section{Simulation studies}
In this section, we carry out simulation studies to evaluate the
performance of the proposed estimator $\hat\beta_{\omega h}$ by
comparing the biases and mean squared errors (MSE) with the
Gehan-estimator $\hat\beta_G$, $\hat\beta_{\omega}$ derived from
$S_{\omega h}(\beta)=0$ with $h_{ik}=h_{jl}=1$, and the smoothed estimator
$\tilde\beta_{\omega h}$  from $\tilde S_{\omega h}(\beta)=0$.

In the simulation studies, we generate the data from model
(\ref{eq:model}) with  $p=2$, and $\beta=(1.2,1.5)^{\rm T}$. Cluster
sizes $n_i$ are sampled from $3$ to $10$ with equal probability. The
censoring times $C_{ik}$ are generated from a uniform distribution
$U(0,\tau)$, where $\tau$ controls the rate of the censoring. The
rates of the censoring are taken as $15\%$ and $30\%$.  The error
terms are generated from a multivariate normal distribution
$\mbox{N}(0,\Sigma(\rho))$ and a multivariate t-distribution
$\mbox{T}(0,\Sigma(\rho))$ with three degrees of freedom, where
$\Sigma(\rho)$ is an exchangeable matrix with parameter $\rho=0.5$
and $0.8$. Note that the correlation coefficient between $T_{ik}$
and $T_{il}$ $(k \neq l)$ is $(e^{\rho}-1)/(e-1)$. The covariate
$X_{ik1}$ is a cluster-level covariate in which $X_{ik1}$ does not
change within each subject or cluster, and independently generated
from the standard normal distribution. The covariate $X_{ik2}$ are
with-cluster covariate in which the covariate varies within each
subject or cluster, and independently generated from the standard
normal distribution. The covariate $X_{ik2}$ is contaminated by
adding an outlier equal to $5$ with a probability of $0$ or $5\%$.
For the case, $1000$ simulations are carried out. The simulation
results are given in Tables \ref{case1}-\ref{case4} present the
biases and mean squared errors for each case. Tables
\ref{casev1}-\ref{casev2} show the empirical variances (Evar) and
the variances (Ivar) using the iterative method of \S 2 for
simultaneously estimating the regression parameters and covariance
matrix.




From Tables \ref{case1}-\ref{case2}, we can see that both biases and MSEs of the
estimates increase as censoring rate increases, and all the
estimates are unbiased when there are no outliers. The mean square errors decrease as the number of cluster increases.
 When there exist
outliers (Tables \ref{case3}-\ref{case4}),  the proposed estimate
$\hat\beta_{\omega h}$  has much smaller biases and MSEs than
$\hat\beta_G$ and $\hat\beta_{\omega}$. The mean squared errors of
the smoothed estimate $\tilde \beta_{\omega h}$ are similar to those
of the nonsmoothed estimate $\hat\beta_{\omega}$ across all cases.
When the covariate is a within-cluster covariate, the estimates
$\hat\beta_G$ corresponding to $\beta_1$ performs better than
$\hat\beta_{\omega}$ and $\hat\beta_{\omega h}$. However, when the
covariate is cluster-level covariate, $\hat\beta_{\omega}$ and
$\hat\beta_{\omega h}$ corresponding to $\beta_1$ perform better
than $\hat\beta_G$. From Tables \ref{casev1}-\ref{casev2}, we can
see that the variance estimates obtained via the iterate method (for
simultaneously estimating the regression parameters and covariance
matrix) for $\beta_1$ and $\beta_2$ are accurate and similar to the
empirical variance  estimates across all simulation studies.
Overall, the results presented in Tables \ref{casev1}-\ref{casev2}
suggest that the smoothing parameter has a minimal impact on the
bias or actual variance of the regression parameter estimates, and
the proposed estimate $\hat\beta_{\omega h}$ are robust and
efficient.

\section{Analysis of real medical data}
In this section, we illustrate the  proposed method by analyzing two
real longitudinal data sets.

The first one is  a longitudinal and survival dataset collected in a
recent clinical trial, which was described by Guo and Carlin (2004).
In this trial, a total of $467$ HIV-infected patients were enrolled
and randomly assigned to receive either didanosine (ddI) or
zalcitabine (ddC). CD4 counts were recorded at study entry and again
at $2$, $6$, $12$, and $18$-month visits, and the times to death
were also recorded. Due to death or censoring of the patients, the
data is unbalanced. For full details regarding the conduct of the
trial the reader is referred to Abrams et al. (1994) and Goldman et
al. (1996). The dataset is available in the JM package in
statistical software R.

\begin{figure}[h]
\begin{center}
\includegraphics[width=80mm,height=80mm]{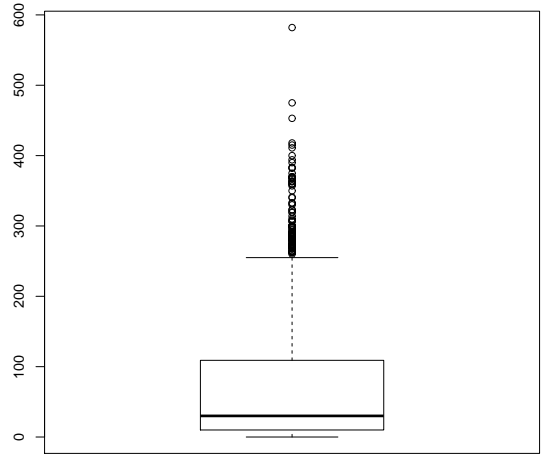}
 \caption{The boxplot of the CD4 values. }
\label{real1fig}       
\end{center}
\end{figure}

In this paper, we are interested in whether the time to death or
censoring of the patients is different for the ddI and ddC groups.
Let $T_{i}$ be the time to death or censoring of the $i$th patient.
We include five covariates as main effects in our analysis: CD4
counts, observation  time  at which the CD4 cells count was recorded
(obstime),  drug ($\mbox{ddI}=1$, $\mbox{ddC}=0$), gender
($\mbox{male}=1$, $\mbox{female}=-1$),  PrevOI (previous
opportunistic infection (AIDS diagnosis) at study entry $= 1$, no
AIDS diagnosis $=-1$), and AZT (AZT failure $= 1$, AZT intolerance
$=-1$). Note that covariates are cluster-level covariates except CD4
and $\mbox{obstime}$. Figure \ref{real1fig} indicates that the CD4
may include some underlying  outliers which are larger than $281$.
We analyze the data by the following AFT model,
$$
\log(T_{i})=\beta_0+\beta_1\mbox{CD}_{ik}+\beta_2\mbox{obstime}_{ik}+\beta_3
\mbox{drug}_i+\beta_4\mbox{gender}_i+\beta_5 \mbox{prevOI}_i+\beta_6
\mbox{AZT}_i+\epsilon_{ik}.
$$
We estimate the parameters by the same method in simulation studies.
Parameter estimates and their standard errors are given in Table
\ref{real1}. We can see that  the estimates obtained from different
methods are similar. Furthermore, $\tilde\beta_{\omega h}$  has
smaller standard errors than $\hat\beta_G$ and $\hat\beta_{\omega}$
for the cluster-level covariates. However, for within-cluster
covariates, CD4 and $\mbox{obstime}$, the standard errors of
$\tilde\beta_{\omega h}$ are larger
 than those of $\hat\beta_G$ and $\hat\beta_{\omega}$, which are
 consistent with the findings in our simulation studies.

\begin{figure}[htp]
\begin{center}
\includegraphics[width=80mm,height=80mm]{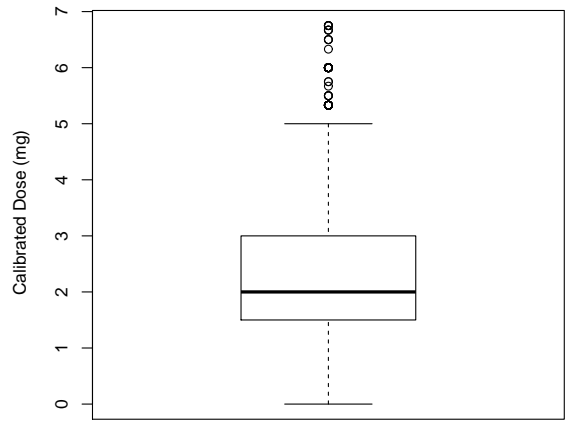}
\caption{The boxplot of the dose in the  SANAD study.
}\label{real2fig}
\end{center}
\end{figure}

The second example is from the standard and new anti-epileptic drugs
(SANAD) study \cite{mar07a,mar07b} with the aim to know whether the
new drug lamotrigine (LTG) is superior to the standard drug
carbamazepine (CBZ) for patients with epilepsy. There were $605$
patients in the trial treated with LTG or CBZ randomly. We consider
the effects of six covariates on the time of drug withdrawal: dose,
treatment (LTG=1, CBZ=0), age, gender (male=1, female=0), and two
indicator variables, with.use (1=withdrawal due to unacceptable
adverse effects, 0=otherwisw) and with.isc(1= withdrawal because of
inadequate seizure control, 0=otherwise). It is noticeable that
these covariates are cluster-level except dose, and there may be
some underlying outliers in dose according to Figure \ref{real2fig}.
We use the following AFT model to analyze the data.
$$
\log(T_{i})=\beta_0+\beta_1\mbox{dose}_{ik}+\beta_2\mbox{treatment}_{i}+\beta_3
\mbox{age}_i+\beta_4\mbox{gender}_i+\beta_5
\mbox{with.uae}_i+\beta_6 \mbox{with.isc}_i+\epsilon_{ik}.
$$
The parameter estimates and their standard errors using different
methods are shown in Table \ref{real2}.

In Table \ref{real2}, it is shown that using different methods
obtains similar estimates. The effect of treatment is positive. In
other words, the conclusion is that LTG is superior to CBZ, which
has been found in \cite{mar07a,mar07b,wil08}.
 Moreover, the standard error of $\tilde{\beta}_{wh}$, the
coefficient of the within-cluster covariate dose, is comparable
with those obtained by  other methods. However, the standard errors
of $\tilde{\beta}_{wh}$ for cluster-level covariates are the
smallest one among the five methods.

\section{Discussion}

The Gehan weight estimating function is monotonic with respect to
regression parameters, is robust against outliers in response, and
has a unique solution.  Therefore, many researchers have utilized it
to estimate parameters in the AFT model (Fygenson and Ritov 1994;
Jin et al. 2006).  However, they did not consider the possible
outliers existing in covariates. Furthermore, the within-cluster
correlation was often ignored for the clustered and censored data.
The proposed method is as simple as the independence model of Jin et
al. (2006), but takes account of within-cluster correlations and
varying cluster sizes. Moreover, it is robust against outliers in
covariates and/or response. When there exist outliers in covariates,
the proposed method  leads to substantial improvement over the
commonly used Gehan estimator. Furthermore, the calculation burden
can be greatly reduced by the induced smoothing method (Brown and
Wang 2005; Johnson and Strawderman 2009).

In this paper, we  only considered the linear regression model. In
fact, the idea can be easily extended to the partial linear model
(Cheng and Wei 2000). The simulation results indicate that the
proposed method depends on the covariate design and  is not fully
efficient; that is  because the correlations are still not well
considered and quantified in this paper. Further work is therefore
needed to incorporate within-cluster correlations into the optimal
parameter estimation.

\section*{\bf Acknowledgements}

~Yan Zhou's research was supported
by the National Natural Science Foundation of China (Grant No.
11701385), the National Statistical Research Project (Grant No.
2017LY56), the Doctor Start Fund of Guangdong Province (Grant No.
2016A030310062). Liya Fu's research was supported by the National
Natural Science Foundation of China (11871390), the Fundamental
Research Funds for the Central Universities (No. xjj2017180), the
Natural Science Basic Research Plan in Shaanxi Province of China
(2018JQ1006) and  the Doctoral Programs Foundation of the Ministry
of Education of China (20120201120053). You-Gan Wang wishes to thank the support from the Australian Research Council grant (DP160104292).

\newpage

\noindent{\bf Appendix A: Proof of Theorem 1}

 {\it Proof:}  Suppose
$G_{ik}(\cdot)$ and $g_{ik}(\cdot)$ are
 distribution and marginal density functions of $\tilde
C_{ik}=\log{C_{ik}}-X_{ik}^{\rm T}\beta_0$ conditional on covariates
$(X_i,X_j)$, respectively. A bar above a distribution denotes a
survival function, such as $\bar F_0(\cdot)=1-F_0(\cdot)$.

We first prove $\sqrt{N}{\rm E}\|\tilde{U}_{\omega
h}(\beta)-U_{\omega h}(\beta)\|=o(1)$. Because
\begin{eqnarray*}
\Delta_{ik}  &=& I(T_{ik}\leq C_{ik}) = I(\log{T_{ik}}-X_{ik}^{\rm T}\beta_0)\leq \log{C_{ik}}-X_{ik}^{\rm T}\beta_0)\\
             &=& I(\epsilon_{ik}\leq \tilde{C_{ik}}),\\
I(e_{ik}-e_{jl}\leq 0) &=& I\left\{\log({T_{ik}\wedge C_{ik}})-X_{ik}^{\rm T}\beta_0-[\log({T_{jl}\wedge C_{jl}})-X_{jl}^{\rm T}\beta_0]+a(\beta)\right\}\\
                       &=& I\left\{\log({T_{ik}\wedge C_{ik}})-X_{ik}^{\rm T}\beta_0-X_{jl}^{\rm T}\beta_0+a(\beta)\leq \log{T_{jl}} \right\}\\
                       &\times& I\left\{\log({T_{ik}\wedge C_{ik}})-X_{ik}^{\rm T}\beta_0-X_{jl}^{\rm T}\beta_0+a(\beta)\leq  C_{jl}
                       \right\},
\end{eqnarray*}
here $a(\beta)=d_{ikjl}^{\rm T}(\beta_0-\beta)$, where
$d_{ikjl}=X_{ik}-X_{jl}$. Let $f_{ikjl}(\cdot)$ be the joint density
function of $(\epsilon_{ik},\epsilon_{jl},\tilde C_{ik},\tilde
C_{jl})$ conditional on the covariates $(X_i,X_j)$. For any $i\neq
j$,
$f_{ikjl}(\Theta)=f_{ik}(\epsilon_{ik})f_{jl}(\epsilon_{jl})g_{ik}(\tilde
C_{ik})g_{jl}(\tilde C_{jl})$, where
$\Theta=(\epsilon_{ik},\epsilon_{jl},\tilde C_{ik},\tilde C_{jl})$.
Therefore,
\begin{eqnarray*}
{\rm E}\{U_{\omega
h}(\beta)\}&=& N^{-2}\sum_{i=1}^N\sum_{j=1}^N\sum_{k=1}^{n_i}\sum_{l=1}^{n_j}\E\left\{ \omega_i\omega_jh_{ik}h_{jl}\Delta_{ik}(X_{ik}-X_{jl})I(e_{ik}-e_{jl}\leq 0)\right\}\\
                &=& N^{-2}\sum_{i=1}^N\sum_{j=1}^N\sum_{k=1}^{n_i}\sum_{l=1}^{n_j}\omega_i\omega_jh_{ik}h_{jl}d_{ikjl}\int_{-\infty}^{+\infty}\bar G_{ik}(u)\bar G_{jl}(u+a(\beta))\bar
                F_{jl}(u+a(\beta))f_{ik}(u)\mathrm{d}u.
\end{eqnarray*}
Define $m_{jl}(s)=\bar G_{jl}(s)f_{jl}(s)+g_{jl}(s)\bar F_{jl}(s)$.
Some tedious algebra leads to
\begin{eqnarray*}
{\rm E}\{\tilde U_{\omega h}(\beta)\}&=& N^{-2}\sum_{i,j=1}^N\sum_{k=1}^{n_i}\sum_{l=1}^{n_j}\E\left\{\omega_i\omega_jh_{ik}h_{jl}(X_{ik}-X_{jl})\Delta_{ik}\Phi\left(\frac{e_{jl}-e_{ik}}{N^{-1/2}r_{ikjl}}\right)\right\}\\
                       &=& N^{-2}\sum_{i,j=1}^N\sum_{k=1}^{n_i}\sum_{l=1}^{n_j}\omega_i\omega_jh_{ik}h_{jl}d_{ikjl}\iint\limits_{\mathbb{R}^2}m_{jl}(s)\Phi\left(\frac{s-u-a(\beta)}{N^{-1/2}r_{ikjl}}\right)\bar G_{ik}(u)f_{ik}(u)\mathrm{d}u\mathrm{d}s\\
                       &=& N^{-2}\sum_{i,j=1}^N\sum_{k=1}^{n_i}\sum_{l=1}^{n_j}\omega_i\omega_jh_{ik}h_{jl}d_{ikjl}\iint\limits_{\mathbb{R}^2}\bar G_{jl}(v_{ikjl})\bar F_{jl}(v_{ikjl})\bar G_{ik}(u)f_{ik}(u)\phi(t)\mathrm{d}t\mathrm{d}u,
\end{eqnarray*}
where $v_{ikjl}=u+a(\beta)+\frac{r_{ikjl}t}{\sqrt{N}}$.
\\
For $\bar G_{jl}(v_{ikjl})$ and $\bar
F_{jl}(v_{ikjl})$, taking a second-order Taylor
series expansion and simple calculation leads to,
$$
{\rm E}\{\tilde U_{\omega h}(\beta)\}={\rm E}\{U_{\omega
h}(\beta)\}+N^{-2}\sum_{i=1}^N\sum_{j=1}^N\sum_{k=1}^{n_i}\sum_{l=1}^{n_j}\omega_i\omega_jh_{ik}h_{jl}d_{ikjl}(Q_1-Q_2+Q_3),
$$
where
$$
Q_1=N^{-1}r^2_{ikjl}\iint\limits_{\mathbb{R}^2}\{\bar
G_{ji}(u+a(\beta))f'_{jl}(u^{**})+g'_{jl}(u^{*})\bar
F_{jl}(u+a(\beta))\}\bar
G_{ik}(u)f_{ik}(u)\phi(t)t^2\mathrm{d}u\mathrm{d}t,
$$
$$
Q_2=N^{-3/2}r_{ikjl}^3\iint\limits_{\mathbb{R}^2}\{g_{jl}(u+a(\beta))f_{jl}(u^{**})+g'_{jl}(u^{*})f_{jl}(u+a(\beta))\}\bar
G_{ik}(u)f_{ik}(u)\phi(t)t^3\mathrm{d}t\mathrm{d}u,
$$ and
$Q_3=N^{-2}r_{ikjl}^4\iint\limits_{\mathbb{R}^2}g'_{jl}(u^*)f'_{jl}(u^{**})\bar
G_{ik}(u)f_{ik}(u)\phi(t)t^4\mathrm{d}t\mathrm{d}u$, where $u^*$ and
$u^{**}$ lie between $u+a(\beta)$ and
$u+a(\beta)+N^{-1/2}r_{ikjl}t$. Therefore,
\begin{eqnarray*}
\sqrt{N}\|{\rm E}\{\tilde U_{\omega h}(\beta)-U_{\omega h}(\beta)\}\|&  =  & N^{-3/2}\|\sum_{i=1}^N\sum_{j=1}^N\sum_{k=1}^{n_i}\sum_{l=1}^{n_j}\omega_i\omega_jh_{ik}h_{jl}d_{ikjl}(Q_1-Q_2+Q_3)\|\\
                                              &\leq & N^{-3/2} \sum_{i=1}^N\sum_{j=1}^N\sum_{k=1}^{n_i}\sum_{l=1}^{n_j}\omega_i\omega_jh_{ik}h_{jl}\|d_{ikjl}\|(Q_1+|Q_2|+Q_3).
\end{eqnarray*}
Under condition C4 and C5, there exist constants $M_1, M_2, M_3$,
which satisfy $Q_1 \leq N^{-1}r^2_{ikjl}M_1$, $Q_2 \leq
N^{-3/2}r_{ikjl}^3\iint\limits_{\mathbb{R}^2}M_2f_{ik}(u)\phi(t)|t^3|\mathrm{d}t\mathrm{d}u=\sqrt{2/\pi}N^{-3/2}r_{ikjl}^3M_2$,
and $Q_3\leq
N^{-2}r_{ikjl}^4\iint\limits_{\mathbb{R}^2}M_3f_{ik}(u)\phi(t)t^4$\\
$\mathrm{d}t\mathrm{d}u=6N^{-2}r_{ikjl}^4M_3$.
Moreover, because $\Gamma=O(1)$, $r_{ikjl}=\sqrt{d_{ikjl}^{\rm T}
\Gamma^2 d_{ikjl}} \leq \sqrt{2}\max_{1\leq i \leq
N}\|X_{ik}\|O(1)$. Therefore, $Q_1 \leq 2M_1\max_{1\leq i \leq
N}\|X_{ik}\|^2O(N^{-1})$, $Q_2 \leq 4M_2\max_{1\leq i \leq
N}\|X_{ik}\|^3O(N^{-3/2})$, and  $Q_3 \leq 12M_3\max_{1\leq i \leq
N}\|X_{ik}\|^4O(N^{-2})$. Under condition C2, we obtain
$\sqrt{N}\|{\rm E}\{\tilde U_{\omega h}(\beta)-U_{\omega
h}(\beta)\}\|=o(1)$. By Chebyshev inequality, $\sqrt{N}\{\tilde
U_{\omega h}(\beta)-U_{\omega h}(\beta)\}=o_p(1)$. $\Box$

\newpage
\noindent{\bf Appendix B: proof of Theorem 2}

\noindent The following lemma is required in order to prove Theorem
2.

{\it Lemma 1.}  Under conditions C1-C5,
$$
\sup_{\beta\in\mathbb{B}}|\tilde L_{\omega h}(\beta)-L_{\omega
h}(\beta)|\xrightarrow {a.s.}0.
$$
{\it Proof.}
\begin{eqnarray*}
|\tilde L_{\omega h}(\beta)-L_{\omega h}(\beta)|&   = &
N^{-2}\left|\sum_{i,j=1}^N\sum_{k=1}^{n_i}\sum_{l=1}^{n_j}
\omega_i\omega_jh_{ik}h_{jl}\Delta_{ik}Z_{ikjl}\right|\\
   &\leq & N^{-2}\sum_{i,j=1}^N\sum_{k=1}^{n_i}\sum_{l=1}^{n_j}
   \omega_i\omega_jh_{ik}h_{jl}\left|\Delta_{ik}Z_{ikjl}\right|\\
                   &=&H_1+H_2,
\end{eqnarray*}
where
\begin{eqnarray*}
Z_{ikjl}=\left\{e_{jlik}\left[\Phi\left(\frac{\sqrt{N}e_{jlik}}{r_{ikjl}}\right)-I(e_{ikjl}\leq0)\right]
  + \frac{1}{\sqrt{N}}r_{ikjl}\phi\left(\frac{{\sqrt{N}}e_{jlik}}{r_{ikjl}}\right)\right\},
\end{eqnarray*}

\begin{eqnarray*}
H_1 &=& N^{-2}\sum_{i=1}^N\sum_{j=1}^N\sum_{k=1}^{n_i}\sum_{l=1}^{n_j}\omega_i\omega_jh_{ik}h_{jl}\left| e_{jlik}\left[\Phi\left(\frac{{\sqrt{N}}e_{jlik}}{r_{ikjl}}\right)-I(e_{ikjl}\leq0)\right]\right|,\\
\end{eqnarray*}
and
$$
H_2=N^{-2}\sum_{i=1}^N\sum_{j=1}^N\sum_{k=1}^{n_i}\sum_{l=1}^{n_j}\omega_i\omega_jh_{ik}h_{jl}
\left|N^{-1/2}r_{ikjl}\phi\left(\frac{{\sqrt{N}}e_{jlik}}{r_{ikjl}}\right)\right|.
$$
Let $t_{ikjl}=e_{ikjl}/(N^{-1/2}r_{ikjl})$, we have
\begin{eqnarray*}
H_1 &=&
N^{-2}\sum_{i=1}^N\sum_{j=1}^N\sum_{k=1}^{n_i}\sum_{l=1}^{n_j}\omega_i\omega_jh_{ik}h_{jl}|N^{-1/2}r_{ikjl}
        t_{ikjl}\{\Phi(-t_{ikjl})-I(t_{ikjl}\leq
        0)\}|\\
    &=& N^{-5/2}\sum_{i=1}^N\sum_{j=1}^N\sum_{k=1}^{n_i}\sum_{l=1}^{n_j}\omega_i\omega_jh_{ik}h_{jl}r_{ikjl}
    t_{ikjl}\Phi(-|t_{ikjl}|)\sgn(t_{ikjl}).
\end{eqnarray*}
Let $t \in \mathbb{R}$, because $\lim_{t\rightarrow \infty}
t\Phi(-|t|)\sgn(t)=0$, hence $t\Phi(-|t|)\sgn(t)$ is bounded. By
condition C5, it follows that $\sup_{\beta\in \mathbb{B}}H_1
\xrightarrow {a.s.}0$, as $N\rightarrow \infty.$ Furthermore,
$|\phi(N^{1/2}e_{jlik}/r_{ikjl})| \leq 1/\sqrt{2\pi}$, thus
$\sup_{\beta\in \mathbb{B}}H_2 \xrightarrow {a.s.}0$, and
$\sup_{\beta\in\mathbb{R}}|L_{\omega h}(\beta)-\tilde L_{\omega
h}(\beta)|\xrightarrow {a.s.}0.$

 {\it \bf Proof of Theorem 2.}
According to Lemma 1 in Johnson and Strawderman (2009), we can get
the similar result under condition C1-C3 as follows:
$$
\sup_{\beta\in\mathbb{B}}| L_{\omega
h}(\beta)-L_0(\beta)|\xrightarrow {a.s.}0.
$$

where
\begin{eqnarray*}
L_0(\beta)&=&\frac{H_{12}+H_{21}}{2}+
\sum_{k=1}^{n_1}\sum_{l=1}^{n_1}\omega^2_1h_{1k}h_{1l}{\rm E}\left\{\Delta_{1k}(e_{1l}-e_{1k})I(e_{1k}-e_{1l}\leq0)\right\},\\
H_{ij}&=&\sum_{k=1}^{n_i}\sum_{l=1}^{n_j}\omega_i\omega_jh_{ik}h_{jl}{\rm
E}\left\{\Delta_{ik}(e_{jl}-e_{ik})I(e_{ik}-e_{jl}\leq0)\right\}.
\end{eqnarray*}
Then, combining Lemma 1 and the triangle inequality
$$
|\tilde L_{\omega h}(\beta)-L_0(\beta)|\leq |\tilde L_{\omega
h}(\beta)-L_{\omega h}(\beta)|+|L_{\omega}(\beta)-L_0(\beta)|,
$$
we obtain $\sup_{\beta\in\mathbb{B}}|\tilde L_{\omega
h}(\beta)-L_{0}(\beta)|\xrightarrow {a.s.}0$. In other words,
$\tilde L_{\omega h}(\beta)$ converges almost surely and uniformly
to the convex function $L_{0}(\beta)$ for $\beta \in \mathbb{B}$. By
condition C6, $L_{0}(\beta)$ is strictly convex at $\beta_0$, and
$\beta_0$ is a unique minimizer of $L_{0}(\beta)$. Therefore,
$\tilde\beta_{\omega h}\xrightarrow {a.s.} \beta_0$.

\newpage

\begin{table}[ht]
\small\centering \caption{Bias and mean squared error (MSE) of the
case that the error terms are generated from a multivariate normal
distribution $\mbox{N}(0,\Sigma(\rho))$. Capital letter C is the
censoring rate.}\label{case1}
\begin{tabular}{lrrrrrrrrrr}
\hline
\multicolumn{ 11}{c}{N=50}     \\
\hline
$\rho$=0.5    &              & \multicolumn{4}{c}{Bias} &&\multicolumn{4}{c}{MSE} \\
\cline{1-1}\cline{3-6}\cline{8-11}
\ \ \\
$\beta$       & C    &$\hat\beta_G$&$\hat\beta_{\omega}$& $\hat\beta_{\omega h}$ & $\tilde{\beta}_{\omega h}$ & &$\hat\beta_G$& $\hat\beta_{\omega}$ &  $\hat\beta_{\omega h}$ & $\tilde{\beta}_{\omega h}$ \\
\hline
$\beta_1=1.2$ &$15\%$& 0.0074     & 0.0075     & 0.0023    & 0.0065     && 0.0155 & 0.0149 & 0.0154 & 0.0156\\

              &$30\%$& 0.0109     & 0.0113     & 0.0027    & 0.0102     && 0.0171 & 0.0164 & 0.0168 & 0.0171 \\

$\beta_2=1.5 $&$15\%$& 0.0062     & 0.0060     & 0.0005    & 0.0059     && 0.0038 & 0.0041 & 0.0041 & 0.0042 \\

              &$30\%$& 0.0103     & 0.0100     & 0.0007    & 0.0101     && 0.0048 & 0.0051 & 0.0051 & 0.0052\\

   \hline
$\rho$=0.8    &              & \multicolumn{4}{c}{Bias} &&\multicolumn{4}{c}{MSE} \\
\cline{1-1}\cline{3-6}\cline{8-11}
\ \ \\
$\beta$       & C     &$\hat\beta_G$ & $\hat\beta_{\omega}$ & $\hat\beta_{\omega h}$ & $\tilde{\beta}_{\omega h}$ & &$\hat\beta_G$   & $\hat\beta_{\omega}$ & $\hat\beta_{\omega h}$ & $\tilde{\beta}_{\omega h}$ \\
\hline
$\beta_1=1.2$ &$15\%$  & 0.0076 & 0.0076  & 0.0024 & 0.0067 && 0.0224 & 0.0209  & 0.0218 & 0.0220 \\
              &$30\%$  & 0.0122 & 0.0126  & 0.0039 & 0.0115 && 0.0246 & 0.0230  & 0.0236 & 0.0240\\
$\beta_2=1.5$ &$15\%$  & 0.0061 & 0.0063  & 0.0009 & 0.0062 && 0.0039 & 0.0042  & 0.0043 & 0.0043\\
              &$30\%$  & 0.0108 & 0.0108  & 0.0014 & 0.0109 && 0.0048 & 0.0052  & 0.0051 & 0.0053\\

\hline
\multicolumn{ 11}{c}{N=100}     \\
\hline
$\rho$=0.5    &              & \multicolumn{4}{c}{Bias} &&\multicolumn{4}{c}{MSE} \\
\cline{1-1}\cline{3-6}\cline{8-11}
\ \ \\
$\beta$       & C          &$\hat\beta_G$ & $\hat\beta_{\omega}$ & $\hat\beta_{\omega h}$ & $\tilde{\beta}_{\omega h}$ & &$\hat\beta_G$& $\hat\beta_{\omega}$ & $\hat\beta_{\omega h}$ & $\tilde{\beta}_{\omega h}$ \\
\hline
$\beta_1=1.2$ &$15\%$  & -0.0004  &-0.0001   & -0.0021   &-0.0000   &&0.0067 & 0.0061  & 0.0063 & 0.0063\\
              &$30\%$  &  0.0013  & 0.0018   & -0.0018   & 0.0019   &&0.0075 & 0.0069  & 0.0070 & 0.0071\\
$\beta_2=1.5 $&$15\%$  &  0.0034  & 0.0036   &  0.0008   & 0.0035   &&0.0018 & 0.0020  & 0.0020 & 0.0020\\
              &$30\%$  &  0.0049  & 0.0049   &  0.0000   & 0.0048   &&0.0024 & 0.0026  & 0.0026 & 0.0026\\

\hline

$\rho$=0.8    &              & \multicolumn{4}{c}{Bias} &&\multicolumn{4}{c}{MSE} \\
\cline{1-1}\cline{3-6}\cline{8-11}
\ \ \\
$\beta$       & C          &$\hat\beta_G$   & $\hat\beta_{\omega}$ &  $\hat\beta_{\omega h}$ & $\tilde{\beta}_{\omega h}$ & &$\hat\beta_G$   & $\hat\beta_{\omega}$ & $\hat\beta_{\omega h}$ & $\tilde{\beta}_{\omega h}$ \\
\hline
$\beta_1=1.2$ &$15\%$  & -0.0000 & 0.0004   & -0.0016  & 0.0005  && 0.0098 & 0.0087  & 0.0090 & 0.0090\\
              &$30\%$  &  0.0020 & 0.0025   & -0.0010  & 0.0027  && 0.0108 & 0.0097  & 0.0099 & 0.0100\\
$\beta_2=1.5 $&$15\%$  &  0.0038 & 0.0041   &  0.0014  & 0.0041  && 0.0020 & 0.0021  & 0.0022 & 0.0022\\
              &$30\%$  &  0.0053 & 0.0055   &  0.0008  & 0.0056  && 0.0026 & 0.0028  & 0.0029 & 0.0029\\
\hline
\end{tabular}
\end{table}
\begin{table}[ht]
 \centering \caption{Bias and mean squared error (MSE) of
the case that the error terms are generated from a multivariate t
distribution with three degrees of freedom
$\mbox{T}_3(0,\Sigma(\rho))$. Capital letter C is the censoring
rate.}\label{case2}
\begin{tabular}{lrrrrrrrrrr}
\\
\hline
\multicolumn{11}{c}{$\mbox{N}=50$}  \\
\hline
$\rho=0.5$    &              & \multicolumn{4}{c}{Bias} & & \multicolumn{4}{c}{MSE}  \\
\cline{1-1}\cline{3-6}\cline{8-11}
\\
$\beta$       & C          &$\hat\beta_G$   & $\hat\beta_{\omega}$  & $\hat\beta_{\omega h}$ & $\tilde{\beta}_{\omega h}$ & & $\hat\beta_G$ & $\hat\beta_{\omega}$ & $\hat\beta_{\omega h}$ & $\tilde{\beta}_{\omega h}$ \\
\hline
$\beta_1=1.2$ &$15\%$&   0.0042 &  0.0053 &   0.0014 &  0.0050 & & 0.0222 &  0.0201 &  0.0199 &  0.0202 \\

              &$30\%$&   0.0103 &  0.0091 &   0.0020 &  0.0089 & & 0.0272 &  0.0247 &  0.0244 &  0.0248\\

$\beta_2=1.5 $&$15\%$&   0.0039 &  0.0036 &  -0.0001 &  0.0040 & & 0.0056 &  0.0062 &  0.0064 &  0.0065\\

              &$30\%$&   0.0131 &  0.0133 &   0.0040 &  0.0122 & & 0.0076 &  0.0084 &  0.0084 &  0.0086\\

   \hline
$\rho=0.8$    &              & \multicolumn{4}{c}{Bias} & & \multicolumn{4}{c}{MSE}  \\
\cline{1-1}\cline{3-6}\cline{8-11}
\\
$\beta$       & C          &$\hat\beta_G$   & $\hat\beta_{\omega}$  & $\hat\beta_{\omega h}$ & $\tilde{\beta}_{\omega h}$ & & $\hat\beta_G$   & $\hat\beta_{\omega}$ &  $\hat\beta_{\omega h}$ & $\tilde{\beta}_{\omega h}$ \\
\hline
$\beta_1=1.2$ &$15\%$&  0.0112 &  0.0092 &  0.0054 &  0.0091 & & 0.0340 &  0.0307 &    0.0303 &  0.0307 \\

              &$30\%$&  0.0167 &  0.0143 &  0.0077 &  0.0144 & & 0.0385 &  0.0344 &    0.0339 &  0.0345 \\

$\beta_2=1.5 $&$15\%$&  0.0062 &  0.0074 &  0.0028 &  0.0070 & & 0.0055 &  0.0064 &    0.0065 &  0.0066\\

              &$30\%$&  0.0134 &  0.0138 &  0.0045 &  0.0124 & & 0.0074 &  0.0082 &    0.0081 &  0.0084\\

   \hline
\multicolumn{11}{c}{$\mbox{N}=100$}  \\
\hline
$\rho=0.5$    &              & \multicolumn{4}{c}{Bias} & & \multicolumn{4}{c}{MSE}  \\
\cline{1-1}\cline{3-6}\cline{8-11}
\\
$\beta$       & C   &$\hat\beta_G$& $\hat\beta_{\omega}$ & $\hat\beta_{\omega h}$ & $\tilde{\beta}_{\omega h}$ & & $\hat\beta_G$   & $\hat\beta_{\omega}$  & $\hat\beta_{\omega h}$ & $\tilde{\beta}_{\omega h}$ \\
\hline
$\beta_1=1.2$ &$15\%$&   0.0072 &  0.0067 &     0.0048 &  0.0066 & & 0.0108 &  0.0101 &    0.0100 &  0.0101 \\

              &$30\%$&   0.0081 &  0.0077 &     0.0045 &  0.0079 & & 0.0126 &  0.0114 &    0.0113 &  0.0114 \\

$\beta_2=1.5 $&$15\%$&   0.0028 &  0.0027 &    -0.0001 &  0.0021 & & 0.0030 &  0.0033 &    0.0033 &  0.0034\\

              &$30\%$&   0.0028 &  0.0030 &    -0.0012 &  0.0029 & & 0.0038 &  0.0042 &    0.0043 &  0.0044\\

   \hline
$\rho=0.8$    &              & \multicolumn{4}{c}{Bias} & & \multicolumn{4}{c}{MSE}  \\
\cline{1-1}\cline{3-6}\cline{8-11}
\\
$\beta$       & C  &$\hat\beta_G$   & $\hat\beta_{\omega}$ &  $\hat\beta_{\omega h}$ & $\tilde{\beta}_{\omega h}$ & & $\hat\beta_G$   & $\hat\beta_{\omega}$ &  $\hat\beta_{\omega h}$ & $\tilde{\beta}_{\omega h}$ \\
\hline
$\beta_1=1.2$ &$15\%$&  0.0033 &  0.0025 &   0.0005 &  0.0024 & & 0.0150 &  0.0138 &   0.0137 &  0.0138 \\

              &$30\%$&  0.0035 &  0.0049 &   0.0017 &  0.0053 & & 0.0191 &  0.0175 &   0.0175 &  0.0176 \\

$\beta_2=1.5 $&$15\%$&  0.0032 &  0.0029 &   0.0001 &  0.0023 & & 0.0031 &  0.0035 &   0.0036 &  0.0036\\

              &$30\%$&  0.0072 &  0.0070 &   0.0031 &  0.0072 & & 0.0039 &  0.0043 &   0.0045 &  0.0045\\

   \hline
\end{tabular}
\end{table}

\begin{table}[ht]
 \centering \caption{Bias and mean squared error (MSE) of
the case that the error terms are generated from a multivariate
normal distribution $\mbox{N}(0,\Sigma(\rho))$ and the covariate
$X_{ik2}$ is contaminated by adding an outlier equal to 5 with a
probability of $5\%$. Capital letter C is the censoring
rate.}\label{case3}
\begin{tabular}{lrrrrrrrrrr}
\\
\hline
\multicolumn{11}{c}{$\mbox{N}=50$}  \\
\hline
$\rho=0.5$    &              & \multicolumn{4}{c}{Bias} & & \multicolumn{4}{c}{MSE}  \\
\cline{1-1}\cline{3-6}\cline{8-11}
\\
$\beta$       & C    &$\hat\beta_G$& $\hat\beta_{\omega}$  & $\hat\beta_{\omega h}$ & $\tilde{\beta}_{\omega h}$ & & $\hat\beta_G$   & $\hat\beta_{\omega}$ &  $\hat\beta_{\omega h}$ & $\tilde{\beta}_{\omega h}$ \\
\hline
$\beta_1=1.2$ &$15\%$&  -0.0085 &  -0.0089 &  -0.0087 &  -0.0043 & & 0.0149 &  0.0141 &    0.0129  & 0.0129 \\

              &$30\%$&  -0.0160 &  -0.0167 &  -0.0142 &  -0.0065 & & 0.0164 &  0.0156 &    0.0143  & 0.0142 \\

$\beta_2=1.5 $&$15\%$&  -0.5184 &  -0.5182 &  -0.0837 &  -0.0814 & & 0.2958 &  0.2972 &    0.0115  & 0.0112\\

              &$30\%$&  -0.5114 &  -0.5107 &  -0.0949 &  -0.0885 & & 0.2898 &  0.2908 &    0.0145  & 0.0134\\

   \hline
$\rho=0.8$    &              & \multicolumn{4}{c}{Bias} & & \multicolumn{4}{c}{MSE}  \\
\cline{1-1}\cline{3-6}\cline{8-11}
\\
$\beta$       & C   &$\hat\beta_G$ & $\hat\beta_{\omega}$  & $\hat\beta_{\omega h}$ & $\tilde{\beta}_{\omega h}$ & & $\hat\beta_G$   & $\hat\beta_{\omega}$  & $\hat\beta_{\omega h}$ & $\tilde{\beta}_{\omega h}$ \\
\hline
$\beta_1=1.2$ &$15\%$&  -0.0091  &  -0.0099  &   -0.0105  & -0.0061 & & 0.0209 &  0.0192 &   0.0183  & 0.0183 \\

              &$30\%$&  -0.0158  &  -0.0168  &   -0.0147  & -0.0068 & & 0.0226 &  0.0210 &   0.0200  & 0.0200 \\

$\beta_2=1.5 $&$15\%$&  -0.5150  &  -0.5147  &   -0.0844  & -0.0820 & & 0.2927 &  0.2946 &   0.0116  & 0.0112\\

              &$30\%$&  -0.5089  &  -0.5082  &   -0.0952  & -0.0887 & & 0.2877 &  0.2890 &   0.0146  & 0.0135\\

   \hline
\multicolumn{11}{c}{$\mbox{N}=100$}  \\
\hline
$\rho=0.5$    &              & \multicolumn{4}{c}{Bias} & & \multicolumn{4}{c}{MSE}  \\
\cline{1-1}\cline{3-6}\cline{8-11}
\\
$\beta$       & C          &$\hat\beta_G$   & $\hat\beta_{\omega}$  & $\hat\beta_{\omega h}$ & $\tilde{\beta}_{\omega h}$ & & $\hat\beta_G$   & $\hat\beta_{\omega}$  & $\hat\beta_{\omega h}$ &$\tilde{\beta}_{\omega h}$ \\
\hline
$\beta_1=1.2$ &$15\%$& -0.0115 &  -0.0119 &    -0.0079 &  -0.0056 & & 0.0082 &  0.0075 &    0.0070 &  0.0070\\

              &$30\%$& -0.0218 &  -0.0221 &    -0.0128 &  -0.0088 & & 0.0091 &  0.0086 &    0.0078 &  0.0078 \\

$\beta_2=1.5 $&$15\%$& -0.5008 &  -0.5006 &    -0.0823 &  -0.0810 & & 0.2634 &  0.2641 &    0.0091 &  0.0089\\

              &$30\%$& -0.5013 &  -0.5007 &    -0.0942 &  -0.0909 & & 0.2646 &  0.2649 &    0.0118 &  0.0112\\

   \hline
$\rho=0.8$    &              & \multicolumn{4}{c}{Bias} & & \multicolumn{4}{c}{MSE}  \\
\cline{1-1}\cline{3-6}\cline{8-11}
\\
$\beta$       & C &$\hat\beta_G$&$\hat\beta_{\omega}$ & $\hat\beta_{\omega h}$ & $\tilde{\beta}_{\omega h}$ & & $\hat\beta_G$   & $\hat\beta_{\omega}$  & $\hat\beta_{\omega h}$ &$\tilde{\beta}_{\omega h}$ \\
\hline
$\beta_1=1.2$ &$15\%$&  -0.0116 &  -0.0122 &  -0.0081 &  -0.0058 & & 0.0115 &  0.0104 &   0.0100 &  0.0100 \\

              &$30\%$&  -0.0231 &  -0.0222 &  -0.0123 &  -0.0082 & & 0.0119 &  0.0111 &   0.0103 &  0.0102 \\

$\beta_2=1.5 $&$15\%$&  -0.4991 &  -0.4993 &  -0.0817 &  -0.0806 & & 0.2620 &  0.2638 &   0.0091 &  0.0089\\

              &$30\%$&  -0.5023 &  -0.5030 &  -0.0911 &  -0.0879 & & 0.2664 &  0.2685 &   0.0114 &  0.0108\\

   \hline
\end{tabular}
\end{table}

\begin{table}[ht]
 \centering \caption{Bias and mean squared error (MSE) of
the case that the error terms are generated from a multivariate t
distribution with three degrees of freedom
$\mbox{T}_3(0,\Sigma(\rho))$ and the covariate $X_{ik2}$ is
contaminated by adding an outlier equal to 5 with a probability of
$5\%$. Capital letter C is the censoring rate.}\label{case4}
\begin{tabular}{lrrrrrrrrrr}
\\
\hline
\multicolumn{11}{c}{$\mbox{N}=50$}  \\
\hline
$\rho=0.5$    &              & \multicolumn{4}{c}{Bias} & & \multicolumn{4}{c}{MSE}  \\
\cline{1-1}\cline{3-6}\cline{8-11}
\\
$\beta$       & C  &$\hat\beta_G$& $\hat\beta_{\omega}$  & $\hat\beta_{\omega h}$ & $\tilde{\beta}_{\omega h}$ & &$\hat\beta_G$& $\hat\beta_{\omega}$  &$\hat\beta_{\omega h}$ &$\tilde{\beta}_{\omega h}$ \\
\hline
$\beta_1=1.2$ &$15\%$&   -0.0076 &  -0.0082 &  -0.0084  &  -0.0051 & & 0.0249 &  0.0240  &  0.0203 &  0.0205 \\

              &$30\%$&   -0.0106 &  -0.0080 &  -0.0070  &   0.0024 & & 0.0272 &  0.0256  &  0.0234 &  0.0238\\

$\beta_2=1.5 $&$15\%$&   -0.5704 &  -0.5721 &  -0.0889  &  -0.0878 & & 0.3533 &  0.3569  &  0.0153 &  0.0152\\

              &$30\%$&   -0.5647 &  -0.5656 &  -0.1083  &  -0.1025 & & 0.3493 &  0.3523  &  0.0208 &  0.0199\\

   \hline
$\rho=0.8$    &              & \multicolumn{4}{c}{Bias} & & \multicolumn{4}{c}{MSE}  \\
\cline{1-1}\cline{3-6}\cline{8-11}
\\
$\beta$       & C  &$\hat\beta_G$& $\hat\beta_{\omega}$ &  $\hat\beta_{\omega h}$ & $\tilde{\beta}_{\omega h}$ & & $\hat\beta_G$   & $\hat\beta_{\omega}$ &  $\hat\beta_{\omega h}$ &$\tilde{\beta}_{\omega h}$ \\
\hline
$\beta_1=1.2$ &$15\%$&   0.0011 &  -0.0004 &     0.0023 &   0.0059 & & 0.0393 &  0.0363 &    0.0325 &  0.0330 \\

              &$30\%$&  -0.0061 &  -0.0061 &    -0.0041 &   0.0026 & & 0.0422 &  0.0384 &    0.0350 &  0.0354 \\

$\beta_2=1.5 $&$15\%$&  -0.5589 &  -0.5577 &    -0.0893 &  -0.0880 & & 0.3421 &  0.3431 &    0.0153 &  0.0151\\

              &$30\%$&  -0.5543 &  -0.5556 &    -0.1005 &  -0.0952 & & 0.3396 &  0.3446 &    0.0196 &  0.0188\\

   \hline
   \ \ \\
\multicolumn{11}{c}{$\mbox{N}=100$}  \\
\hline
$\rho=0.5$    &              & \multicolumn{4}{c}{Bias} & & \multicolumn{4}{c}{MSE}  \\
\cline{1-1}\cline{3-6}\cline{8-11}
\\
$\beta$       & C     &$\hat\beta_G$ & $\hat\beta_{\omega}$  & $\hat\beta_{\omega h}$ & $\tilde{\beta}_{\omega h}$ & & $\hat\beta_G$& $\hat\beta_{\omega}$ & $\hat\beta_{\omega h}$ & $\tilde{\beta}_{\omega h}$ \\
\hline
$\beta_1=1.2$ &$15\%$&   -0.0016 &  -0.0013 &     0.0013 &   0.0032 & & 0.0129 &  0.0121 &    0.0105 &  0.0106 \\

              &$30\%$&   -0.0173 &  -0.0166 &    -0.0098 &  -0.0062 & & 0.0132 &  0.0123 &    0.0111 &  0.0111 \\

$\beta_2=1.5 $&$15\%$&   -0.5785 &  -0.5792 &    -0.0938 &  -0.0932 & & 0.3516 &  0.3531 &    0.0126 &  0.0125\\

              &$30\%$&   -0.5615 &  -0.5618 &    -0.1039 &  -0.1011 & & 0.3314 &  0.3331 &    0.0153 &  0.0148\\

   \hline
$\rho=0.8$    &              & \multicolumn{4}{c}{Bias} & & \multicolumn{4}{c}{MSE}  \\
\cline{1-1}\cline{3-6}\cline{8-11}
\\
$\beta$       & C &$\hat\beta_G$& $\hat\beta_{\omega}$ & $\hat\beta_{\omega h}$ & $\tilde{\beta}_{\omega h}$ & &$\hat\beta_G$& $\hat\beta_{\omega}$& $\hat\beta_{\omega h}$ &$\tilde{\beta}_{\omega h}$ \\
\hline
$\beta_1=1.2$ &$15\%$&  -0.0014 &  -0.0030 &    -0.0011 &   0.0007 & & 0.0186 &  0.0166 &    0.0151 &  0.0152 \\

              &$30\%$&  -0.0168 &  -0.0164 &    -0.0108 &  -0.0072 & & 0.0186 &  0.0167 &    0.0151 &  0.0152 \\

$\beta_2=1.5 $&$15\%$&  -0.5738 &  -0.5747 &    -0.0947 &  -0.0940 & & 0.3469 &  0.3495 &    0.0130 &  0.0129\\

              &$30\%$&  -0.5650 &  -0.5674 &    -0.1045 &  -0.1019 & & 0.3382 &  0.3422 &    0.0159 &  0.0154\\

   \hline
\end{tabular}
\end{table}


\begin{table}[ht]
\centering \caption{Evar and Ivar correspond to the empirical
variance  and the variance of the estimator $\tilde\beta_{wh}$ using
the iterative method of \S 2 for simultaneously estimating the
regression parameters and covariance matrix. The error terms are
generated from a multivariate normal distribution
$\mbox{N}(0,\Sigma(\rho))$ and a multivariate t distribution
$\mbox{T}_3(0,\Sigma(\rho))$. }\label{casev1}
\begin{tabular}{lrrrrrrrrrrrr}
 \hline
\multicolumn{ 13}{c}{~~~~~~~~~Error terms $(\epsilon_{i1},\cdots,\epsilon_{in})\sim \mbox{N}(0,\Sigma(\rho))$~~~~~~~~~~ }     \\
\hline
             &       & \multicolumn{ 5}{c}{N=50}            & & \multicolumn{ 5}{c}{N=100} \\
\cline{3-7}\cline{9-13}
 \ \ \\
             &       & \multicolumn{2}{c}{$\hat\beta_1$}   & &\multicolumn{2}{c}{$\hat\beta_2$}&  & \multicolumn{2}{c}{$\hat\beta_1$} & &\multicolumn{2}{c}{$\hat\beta_2$}  \\
\cline{3-4}\cline{6-7}\cline{9-10}\cline{12-13}
             & C     & Ivar   &  Evar   && Ivar   &Evar     & &  Ivar & Evar   &&Ivar    &Evar\\
\hline
$\rho$=0.5  &$15\%$   & 0.0152 &0.0155  &&  0.0036 & 0.0041 & &0.0073 & 0.0063 && 0.0018 & 0.0020   \\
            &$30\%$   & 0.0159 &0.0170  &&  0.0044 & 0.0051 & &0.0077 & 0.0071 && 0.0023 & 0.0026   \\
$\rho=0.8$  &$15\%$   & 0.0196 &0.0220  &&  0.0038 & 0.0043 & &0.0093 & 0.0090 && 0.0019 & 0.0022   \\
            &$30\%$   & 0.0208 &0.0238  &&  0.0047 & 0.0052 & &0.0101 & 0.0100 && 0.0024 & 0.0029   \\
\hline

 \multicolumn{ 13}{c}{~~~~~~~~Error terms $(\epsilon_{i1},\cdots,\epsilon_{in})\sim \mbox{T}_3(0,\Sigma(\rho))$~~~~~~~ }     \\
\hline
             &       & \multicolumn{ 5}{c}{N=50}            & & \multicolumn{ 5}{c}{N=100} \\
\cline{3-7}\cline{9-13}
 \ \ \\
 &     & \multicolumn{2}{c}{$\hat\beta_1$} & &\multicolumn{2}{c}{$\hat\beta_2$} &   & \multicolumn{2}{c}{$\hat\beta_1$} & &\multicolumn{2}{c}{$\hat\beta_2$}  \\
\cline{3-4}\cline{6-7}\cline{9-10}\cline{12-13}
              & C     & Ivar   &  Evar   && Ivar   &Evar      & &  Ivar    & Evar   &&Ivar    &Evar\\
\hline
$\rho=0.5$    &$15\%$  &0.0218 &0.0202 & &0.0053 &0.0065 &  &0.0104 &0.0100 & &0.0027 &0.0034   \\

              &$30\%$  &0.0255 &0.0248 & &0.0071 &0.0085 &  &0.0118 &0.0114 & &0.0038 &0.0044  \\

$\rho=0.8$     &$15\%$  &0.0292 &0.0306 & &0.0057 &0.0066 &  &0.0136 &0.0138 & &0.0030 &0.0036   \\

               &$30\%$  &0.0331 &0.0343 & &0.0075 &0.0083 &  &0.0155 &0.0176 & &0.0039 &0.0045  \\
   \hline
\end{tabular}
\end{table}

\begin{table}[ht]
\scriptsize \centering \caption{ Evar and Ivar correspond to the
empirical variance and the variance of the estimator
$\tilde{\beta}_{wh}$ using the iterative method of $\S2$ for
simultaneously estimating the regression parameters and covariance
matrix. The covariate $X_{ik2}$ is contaminated by adding an outlier
equal to $5$ with a probability of $5\%$. The error terms are
generated from a multivariate normal distribution
$\mbox{N}(0,\Sigma(\rho))$ and a multivariate t distribution
$\mbox{T}_3(0,\Sigma(\rho))$.} \label{casev2}
\begin{tabular}{lrrrrrrrrrrrr}
\\
 \hline
\multicolumn{ 13}{c}{Error terms ($\epsilon_{i1},\cdots,\epsilon_{in}) \sim \mbox{N}(0,\Sigma(\rho))$}     \\
\hline
& & \multicolumn{5}{c}{$\mbox{N}=50$} & &  \multicolumn{5}{c}{$\mbox{N}=100$} \\
\cline{3-7}\cline{9-13}
\\
& & \multicolumn{2}{c}{$\hat{\beta}_1$} & &
\multicolumn{2}{c}{$\hat{\beta}_2$}  & &
 \multicolumn{2}{c}{$\hat{\beta}_1$} & &  \multicolumn{2}{c}{$\hat{\beta}_2$}   \\
\cline{3-4}\cline{6-7}\cline{9-10}\cline{12-13}
       & C     & Ivar & Evar & & Ivar & Evar & & Ivar & Evar & & Ivar & Evar \\
\hline
$\rho=0.5$    &$15\%$   &0.0136 &0.0129 & &0.0040 &0.0046 &  &0.0065 &0.0070 & &0.0020 &0.0023   \\

              &$30\%$   &0.0143 &0.0142 & &0.0050 &0.0055 &  &0.0069 &0.0077 & &0.0026 &0.0030  \\

$\rho=0.8$     &$15\%$  &0.0181 &0.0182 & &0.0042 &0.0045 &  &0.0085 &0.0100 & &0.0022 &0.0025   \\

               &$30\%$  &0.0191 &0.0199 & &0.0052 &0.0056 &  &0.0093 &0.0101 & &0.0027 &0.0031  \\
\hline
\multicolumn{ 13}{c}{Error terms ($\epsilon_{i1},\cdots,\epsilon_{in}) \sim \mbox{T}_3(0,\Sigma(\rho))$}     \\
\hline
& & \multicolumn{5}{c}{$\mbox{N}=50$} & &  \multicolumn{5}{c}{$\mbox{N}=100$} \\
\cline{3-7}\cline{9-13}
\\
& & \multicolumn{2}{c}{$\hat{\beta}_1$} & &
\multicolumn{2}{c}{$\hat{\beta}_2$}  & &
 \multicolumn{2}{c}{$\hat{\beta}_1$} & &  \multicolumn{2}{c}{$\hat{\beta}_2$}   \\
 \cline{3-4}\cline{6-7}\cline{9-10}\cline{12-13}
       & C     & Ivar & Evar & & Ivar & Evar & & Ivar & Evar & & Ivar & Evar \\
\hline

$\rho=0.5$    &$15\%$  &0.0213 &0.0205 & &0.0063 &0.0074 &  &0.0102 &0.0106 & &0.0031 &0.0038   \\

              &$30\%$  &0.0236 &0.0238 & &0.0081 &0.0094 &  &0.0112 &0.0111 & &0.0039 &0.0046  \\

$\rho=0.8$    &$15\%$  &0.0289 &0.0329 & &0.0063 &0.0074 &  &0.0132 &0.0152 & &0.0036 &0.0041   \\

              &$30\%$  &0.0311 &0.0354 & &0.0086 &0.0097 &  &0.0150 &0.0152 & &0.0042 &0.0050  \\
\hline

\end{tabular}
\end{table}


\begin{table}[ht]
\centering \caption{The  estimates and their standard errors (SE) of
the coefficients in the AFT model for the HIV data. }\label{real1}
\begin{tabular}{rrrrrrrrr}
\hline
    Method      &CD4   &obstime & drug   &gender  & prevOI & AZT   \\
 \hline

$\hat\beta_G$            & 0.0050 & 0.0981 &-0.1330 & 0.1051& -0.1977& -0.0053\\
(SE)                       & (0.0068) & (0.0215) & (0.1862) & (0.2196)&  (0.1641)&  (0.0893)\\
$\hat\beta_{\omega}$     & 0.0055 & 0.1215 & -0.1600 & 0.1432& -0.2271& -0.0129 \\
(SE)                       & (0.0066) & (0.0219) & (0.1617) & (0.1682)&  (0.1548)&  (0.0774)\\
$\hat\beta_{\omega h}$   & 0.0090 & 0.1285 &-0.1436 & 0.1596& -0.2579& -0.0302\\
  (SE)                       & (0.0050) & (0.0180) & (0.1246) & (0.1255)&  (0.1211)&  (0.0666)\\
 \hline
 \end{tabular}
\end{table}

\begin{table}[ht]
\centering \caption{The  estimates and their standard errors (SE) of
the coefficients in the AFT model for the SANAD data. }\label{real2}
\begin{tabular}{rrrrrrrrr}
\hline
    Method      &dose   &treatment & age   &gender  & with.uae & with.isc   \\
 \hline

$\hat\beta_I$           & 0.1914  & 0.0055  & 0.0101  & 0.1307   & -4.9838    & -4.4585\\
(SE)                    &(0.0540) & (0.1777)& (0.0050)& (0.1723) &  (0.3901)  &  (0.4035)\\
$\hat\beta_w$           & 0.3034  & 0.1575  & 0.0059  & 0.1058   & -4.6099    & -4.1111\\
(SE)                    & (0.0600)& (0.1727)& (0.0049)& (0.1637) &  (0.3402)  &  (0.3590)\\
 $\hat\beta_{wh}$       & 0.3245  & 0.2217  & 0.0051  & 0.1461   & -3.3281    & -2.8242\\
(SE)                    & (0.0710)& (0.1553)& (0.0048)& (0.1513) &  (0.3307)  &  (0.3498)\\
 \hline
 \end{tabular}
\end{table}

\end{document}